# Unveiling the Role of Electron-Phonon Scattering in Dephasing High-Order Harmonics in Solids

Viacheslav Korolev[1,†], Thomas Lettau[2,†], Vipin Krishna[3], Alexander Croy[4], Michael Zürch[5,6], Christian Spielmann[1,7,8], Maria Wächtler[9], Ulf Peschel[2,8], Stefanie Gräfe[4,8], Giancarlo Soavi[3,8] and Daniil Kartashov[1,8*]

[1]Institute of Optics and Quantum Electronics, Friedrich-Schiller University Jena, Max-Wien-Platz 1, 07743 Jena, Germany

[2]Institute of Condensed Matter Theory and Solid State Optics, Friedrich Schiller University Jena, Fröbelstieg 1, 07743 Jena, Germany

[3]Institute of Solid State Physics, Friedrich-Schiller University Jena, Helmholtzweg 3, 07743 Jena, Germany

[4]Institute of Physical Chemistry, Friedrich Schiller University Jena, Helmholtzweg 4, 07743 Jena, Germany

[5]Department of Chemistry, University of California at Berkeley, Berkeley, California 94720, USA

[6]Materials Sciences Division, Lawrence Berkeley National Laboratory, Berkeley, California 94720, USA

[7]Helmholtz-Institut Jena, Helmholtzweg 4, 07743 Jena, Germany

[8]Abbe Center of Photonics, Albert-Einstein-Straße 6, 07745 Jena, Germany

[9]Chemistry Department and State Research Center OPTIMAS, RPTU Kaiserslautern-Landau, Erwin-Schrödinger-Straße 52, 67663 Kaiserslautern

## Abstract

**High-order harmonic generation (HHG) in solids is profoundly influenced by the dephasing of the coherent electron-hole motion driven by an external laser field. The exact physical mechanisms underlying this dephasing, crucial for accurately understanding and modelling HHG spectra, have remained elusive and controversial, often regarded more as an empirical observation than a firmly established principle. In this work, we present comprehensive experimental findings on the wavelength-dependency of HHG in both single-atomic-layer and bulk semiconductors. These findings are further corroborated by rigorous numerical simulations, employing ab initio real-time, real-space time-dependent density functional theory and semiconductor Bloch equations. Our experimental observations necessitate the introduction of a novel concept: a momentum-dependent dephasing time in HHG. Through detailed analysis, we pinpoint momentum-dependent electron-phonon scattering as the predominant mechanism driving dephasing. This insight significantly advances the understanding of dephasing phenomena in solids, addressing a long-standing debate in the field. Furthermore, our findings pave the way for a novel, all-optical measurement technique to determine electron-phonon scattering rates and establish fundamental limits to the efficiency of HHG in condensed matter.**

The nonlocal nature of high-order harmonic generation (HHG) in both, real coordinate and energy-crystal momentum space, within solids paves the way to new methods of ultrafast time-resolved spectroscopy in condensed matter. According to the well-established physical picture of HHG in solids, the strong field that excites electrons in the conduction band drives them on the half-optical-cycle time scale over many lattice units in real space and up to the entire Brillouin zone in the energy-momentum domain[1-3]. Correspondingly, two nonlinear sources of harmonic generation are at play: intraband currents (due to the non-parabolicity of the band structure) and interband electron-hole recombination, the solid-state analogue of the HHG mechanism in gases[1,2]. The motion of electrons over the entire Brillouin zone further implies that HHG is a powerful probe of the band structure in solids, including specific features such as Berry's curvature, valleys and topological properties that remains encoded in the spectral and polarisation properties of the generated harmonics[4-14].

In contrast to gases, HHG in solids is intrinsically a multiparticle effect. Therefore, the interband (Corkum) mechanism of HHG is extremely sensitive to particle interactions that cause a dephasing of the otherwise phase-locked electron-hole motion. An associated dephasing time was introduced in the theory of HHG in solids as an empirical parameter and serves the purpose of converting rather stochastic emission spectra, calculated within semiconductor Bloch equations (SBE) or density matrix approaches, into the regular harmonic structure observed in experiments[1,15,16]. Although the physical origin of the dephasing was immediately attributed to electron-electron or electron-lattice scattering processes, the value of the dephasing time was somehow linked to the optical cycle[1,16]. Up to date, there is no clear physical assignment on the value of the dephasing time, and the empiric ultrashort duration 2-5 fs, commonly used in simulations, cannot be justified based on the spectroscopic data available from weak field interactions. Contrary, in the regime of perturbative nonlinear optics much longer dephasing times are required to get an agreement with the experiment[17]. At the same time, it was suggested that the concept of dephasing in HHG is artificial and can be omitted when propagation effects are taken properly into account in the HHG process[18]. Thus, the concept of dephasing time in HHG in solids is rather contradictive.

In addition to its strong influence on the characteristics of the emitted spectrum, dephasing plays a decisive role also in the dependence of the harmonic yield on the driving laser wavelength[16,19,20]. In particular, it was shown that the smaller the value of the dephasing time, the higher the exponential index in the drop of HHG efficiency for increasing laser wavelength[19]. Also, it was found experimentally that the efficiency of HHG in solids drops much more abruptly with increasing laser wavelength compared to HHG in gases[20]. Thus, the "reality" of dephasing and the value of the associated dephasing time appear to be directly related to the scaling of the harmonic yield with the laser wavelength. Overall, a deeper understanding of the physical origin, connected with an exact value of the dephasing time, that is needed to model HHG in solids, is one of the most important and long-lasting questions of the field[21,22].

Here, we report on the results of combined experimental and theoretical study to measure and interpret the dephasing time of solid-state systems from the wavelength dependence of HHG spectra at a fixed laser intensity. To circumvent the influence of linear and nonlinear propagation effects, the experiments are first carried out on monolayer $WS_2$ and $MoS_2$ transition metal dichalcogenides (TMDs). Further, we compare the results of atomically thin semiconductors to HHG obtained from bulk polycrystalline CdSe film. Both atomically thin and bulk semiconductors display an ultra-short dephasing time below 10 fs, that we attribute to the highly dispersive momentum dependent electron-phonon scattering. The comparison and similarities between the wavelength dependence of HHG in atomically thin and bulk semiconductors suggest that this momentum dependent dephasing time is a universal feature of solids, rather than a sample-dependent parameter. Thus, besides their fundamental scientific relevance, our findings are crucial for the design of compact attosecond sources based on HHG in solids.

## Results

**High harmonic spectra of atomically thin and bulk semiconductors.** We excite both monolayer ($MoS_2$ and $WS_2$) and bulk (CdSe) semiconductors with intense (up to 2 TW/cm$^2$) femtosecond mid-IR pulses at different wavelengths in the range 3-5 µm and measure high-order harmonic spectra within the 0.2-1 µm spectral region (see Methods and experimental details in Supplementary Fig. S1). In our analysis we focus on harmonics with photon energies larger than the electronic bandgap, namely those that are generated predominantly by the recombination or interband mechanism. TMD monolayers were obtained by mechanical exfoliation and transferred onto a c-cut sapphire substrate (see Supplementary Note 1). The bulk, polycristalline, 140 nm thick, wurtzite symmetry CdSe film was prepared directly on the sapphire substrate *via* chemical bath deposition.

Fig.1 shows the high harmonic spectra obtained from two different TMD monolayers (Panels a-b), oriented along the zigzag direction relative to the laser polarisation, and from the CdSe film at approximately the same 1.5 TW/cm$^2$ peak intensity. The dashed line indicates the electronic bandgap and the gray shaded area represents the integration region that we used to calculate the total HHG yield. In TMD monolayers, both even and odd harmonics are generated up to 5 eV and they exhibit a clear plateau behaviour with a cut-off near 4 eV. In contrast, harmonic spectra measured from the bulk CdSe contain only the odd orders and the intensities exponentially decay with increasing harmonics order. This can be explained by strong re-absorption of photons with energies above the bandgap (1.64 eV) of the material[23]. It is noteworthy that for a thickness of 140 nm, CdSe film dispersion effects like phase and group velocity mismatch are negligible in the entire spectral range of harmonic measurements.

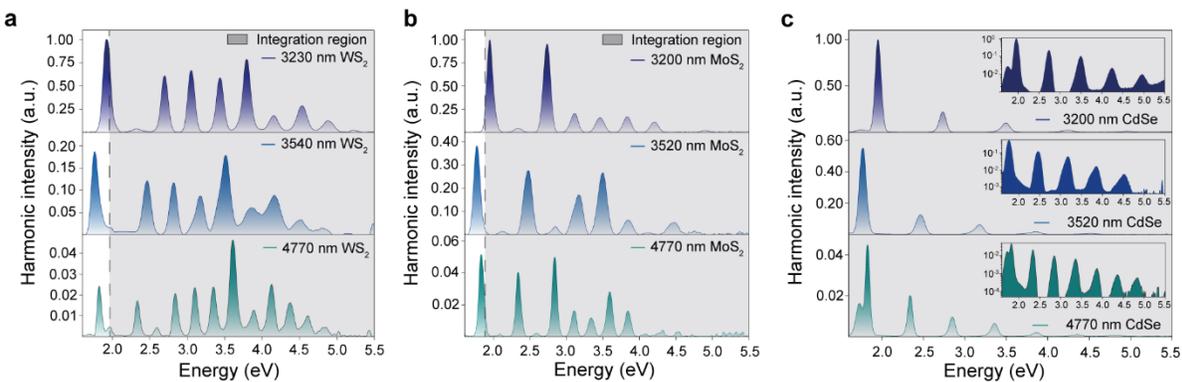

**Fig. 1 Wavelength dependence in solid-state HHG**. HHG spectra on linear scale, corrected to the spectrometer and the detector sensitivity, for three selected laser wavelengths from (**a**) $WS_2$ along zigzag direction, (**b**) $MoS_2$ along zigzag direction and (**c**) CdSe polycrystalline film, inset figures are the same, but in logarithmic scale for better visibility. The shaded area on all figures marks harmonics with energies of quanta above the bandgap and represents the integration region.

**Experimental wavelength dependence.** Fig. 2a shows the experimentally measured wavelength dependence from a monolayer $WS_2$ along the two main symmetry axes: zigzag (top figure) and armchair directions (bottom figure). The blue dots represent the normalised harmonic yield which accounts only for harmonics above the band gap (as discussed in the previous section), measured at different pump laser wavelengths but at a fixed laser intensity of 1.45 TW/cm$^2$ (for details on the intensity calibration see Supplementary Note 2). It is convenient to present such dependencies on a double logarithmic scale: in this way, a power-law decay can be fitted with a linear function, characterising the exponent of its decay by the value of the slope. For monolayer $WS_2$ we find values of such slope of 8.9 ± 1 and 8.2 ± 0.9 along the zigzag and armchair directions respectively (here and further in the text, the absolute value of the slope is given). Similarly, Fig. 2b shows the wavelength dependence of the total HHG yield from a $MoS_2$ monolayer along the zigzag and armchair directions and the CdSe film. The retrieved slopes are 6.6 ± 2 for $MoS_2$ zigzag and 6.2 ± 1.4 for both $MoS_2$ (armchair) and CdSe. In all cases the harmonic yield does not decay monotonically, but instead it shows an oscillatory behaviour. These oscillations might be attributed to a quantum path interference predicted theoretically for HHG in gases[24] and solids[19]. In general, the exponent in the decay of the harmonic yield with increasing laser wavelength, retrieved from the experimental measurements, is significantly larger than in the $\lambda^{-4}$ dependence expected under the assumption of an infinite dephasing time[19]. Thus, our results indicate that dephasing processes, responsible for decoherence in phase-locked electron-hole dynamics, play a crucial role in the process of HHG in solids. Moreover, as we show in the next section, a direct comparison between the numerical simulations by Liu et al. (Ref.19) and our own model suggests that the effective dephasing times in all our samples lie in the range $\approx$5-10 fs.

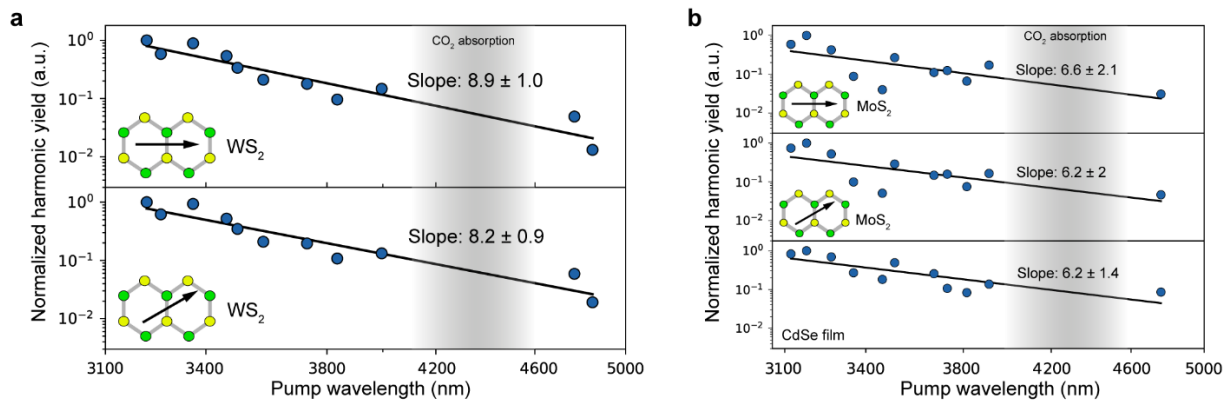

**Fig. 2 Wavelength dependence of the high harmonic yield. a** Normalised harmonic yield as a function of the pump wavelength plotted in a double logarithmic scale from a monolayer $WS_2$ along the zigzag (top figure) and the armchair (bottom figure) directions. **b** The same for $MoS_2$, top figure along the zigzag direction, middle along the armchair and bottom for CdSe film. The black line shows a linear fit and the absolute values of the slopes are shown. The shaded grey area represents the CO2 absorption region where measurements are not possible under ambient conditions. The laser intensity was fixed to 1.45 TW/cm$^2$.

**Scattering processes and k-dependent dephasing time.** To further elucidate the role of the dephasing time in HHG of solids, we carried out real-space and real-time time-dependent density functional (rt-TDDFT) calculations for the wavelength dependence of the HHG yield in an WS$_2$ monolayer, mimicking the experiments (see Methods). The simulations results predict a slope of the decay in the harmonic yield with increasing laser wavelength $\approx$3.3$\pm$0.7 (i.e. $\approx\lambda^{-3.3}$ dependence, see Supplementary Figure 4). Since rt-TDDFT calculations do not involve any dephasing processes, this result already points to the importance and impact of dephasing mechanisms to properly model the experiment. However, before we include the dephasing times in our numerical model, it is worth discussing the physical origin of possible ultrafast dephasing processes occurring in solids (Fig. 3). Here, we stress that after the initial step of strong field excitation *via* electron tunnelling from the valence to the conduction band, the electron-hole motion is non-local both in the band and real space (Fig. 3a). In contrast to nonlinear perturbative and parametric optical spectroscopy of TMDs[25], when after excitation the electron and hole remain in the vicinity of K/K′ valley (or Γ-point in other semiconductors and dielectrics), in HHG the electron-hole motion within a single optical cycle time scale covers a large fraction or even the entire Brillouin zone. Indeed, the amplitude of the lattice momentum acquired by an electron in the laser field can be calculated as $k_f[nm^1] = \frac{eE_0}{\hbar\omega} \approx 2.177\sqrt{I\left[\frac{TW}{cm^2}\right]}\lambda[\mu m]$ or, normalizing to the lattice constant of WS$_2$ $a = 3.188$ Å, $\widetilde{k_f} = \frac{k_f a}{2\pi} \approx 0.1\sqrt{I\left[\frac{TW}{cm^2}\right]}\lambda[\mu m]$. For the shortest wavelength 3.1 μm and the intensity 1.45 TW/cm$^2$ we estimate $\widetilde{k_f} \approx 0.36$. When comparing it to the distances KM$\approx$0.33 and KΓ$\approx$0.67 in momentum space for the motion in the band along the zigzag direction, we conclude that the electrons, excited at K point, role over the border of BZ within one optical half-cycle and cover at least half of the BZ within the opposite polarity half-cycle.

The first collisional process, carrier multiplication (Fig. 3b), is much more efficient in 2D materials than in bulk semiconductors due to quantum confinement[26,27,28]. However, this process becomes efficient only if the excited electron has an excess energy equal or larger than the bandgap (*i.e.*, a total energy larger than twice the bandgap), which is possible by single or several photon absorption from a light source with large photon energy. Under the conditions of strong field excitation with photon energies much smaller than the bandgap, as it is realised in our experiments, the electron tunnels to the bottom of the first conduction band. In this case, carrier multiplication in TMDs is extremely inefficient and can be neglected. This was confirmed in experiments on transport phenomena under DC voltage suggesting that carrier multiplication in this case is primarily due to the carrier contribution from trap defects states[29,30].

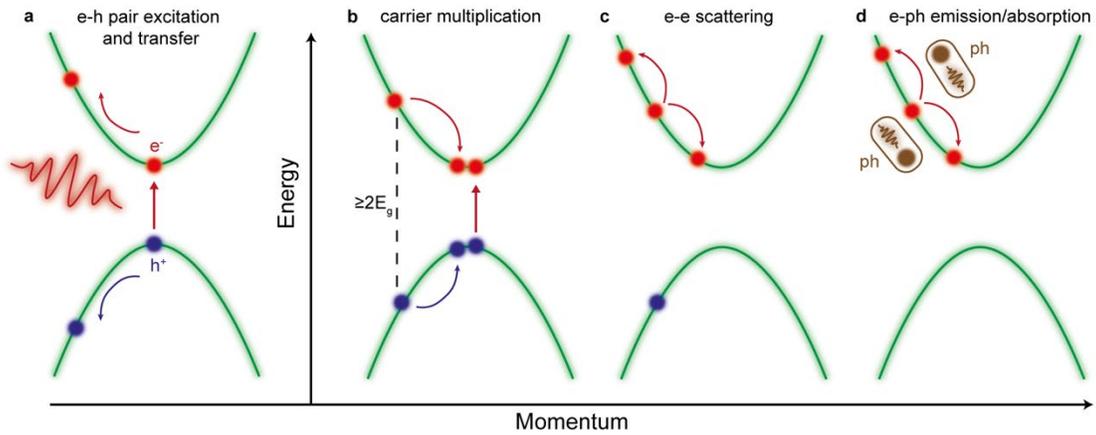

**Fig. 3** Scattering processes in solids. **a** Sketch of tunnel excitation and an electron-hole pair creation that is subsequently accelerated by the laser field, **b** carrier multiplication process, **c** electron-electron and **d** electron-phonon scattering. For the last scattering process both possibilities, phonon absorption and emission, are depicted.

The second scattering process that can contribute to the dephasing in HHG is electron-electron scattering (Fig. 3c). To study the contribution of this process, we analyse the experimental wavelength dependence of the high harmonic yield in monolayer $WS_2$ for different intensities in the driving laser field (Fig. 4). The slope in the wavelength dependence of the harmonic yield undergoes only minor changes (less than 10%) when the laser intensity is changed by more than 70%. However, based on the Keldysh strong-field excitation rate in semiconductors[31,32], we estimate that the carrier density increases by approximately up to one order of magnitude for the intensity variation in that range. Recent experiments on influence of photo-doping on HHG in 2D materials have shown that, under conditions of high density of the excited carriers, when screening effects can be neglected and electron-electron collisions play the major role, the dephasing time scales with the electron density as $\sim 1/N_e$[33]. According to our numerical simulations (see Fig.6 and details further in the text), such variation in the dephasing time would be reflected in a significant change of the slope, which however, is not observed in our experiments (Fig.4). Thus, we conclude that electron-electron scattering plays a minor role in the dephasing responsible for the steep slope in the dependence of the harmonic yield on the driving wavelength.

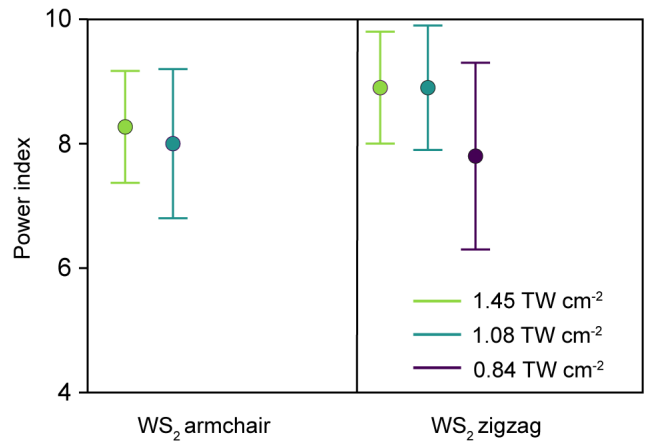

**Fig. 4 Power dependent slope of HHG vs excitation wavelength.** Experimentally measured dependence of the slope in the wavelength-dependent decay of the harmonic yield on the driving laser intensity, and for different crystal orientations.

Finally, we consider the contribution of electron-phonon scattering (Fig. 3d). Using DFT, we calculate the total k-dependent, band-specific electron-phonon scattering rate in $WS_2$, summed over the scattering rates for each phonon mode, populated according to the Boltzmann distribution at room temperature, and including both, phonon absorption and phonon emission, processes (see Supplementary Note 3), as shown in Fig. 5. The striking feature of the k-dependent scattering time, defined as the inverse of the scattering rate, is its sharp dependence on the electron lattice momentum. In the first conduction band it drops abruptly from ≈200 fs at the K-points to below 10 fs as soon as the electron leaves the valley and reduces to ≈2 fs in the vicinity of the Γ-point (see Fig. 5a). A similar ultrashort time scale is found in the valence band for the hole-phonon scattering (Fig. 5b). Our results agree with previous calculations on the electron-phonon scattering contribution to transport phenomena in TMD[34,35]. Fig. 5a also suggests that the average scattering time for electron motion along the armchair direction is slightly longer compared to that along the zigzag direction. This results in smaller values of the slope, as it is retrieved from the experimental measurements (Fig. 2a). Further, numerical simulations performed for a lattice temperature of 20 °K show that the scattering time increases in the

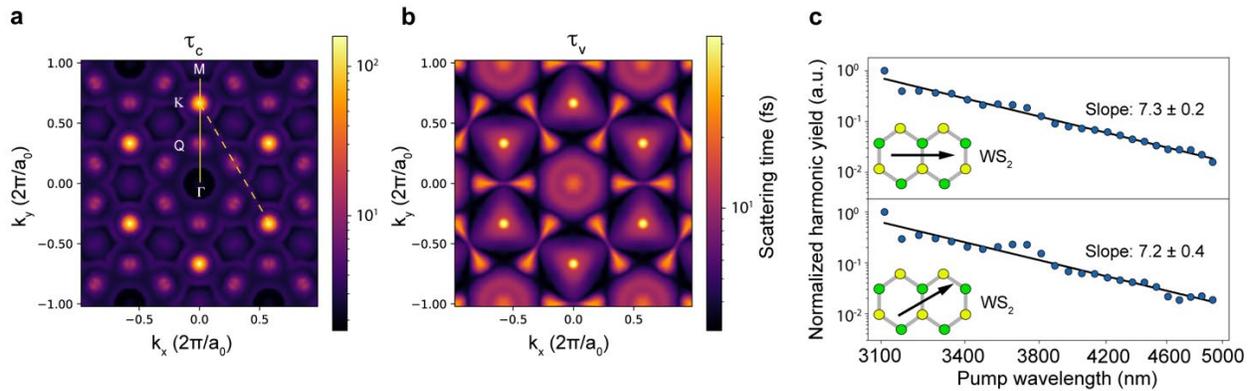

**Fig. 5 Numerical simulations for the k-dependent electron-phonon scattering time.** Two-dimensional map of **a** electron-phonon scattering time in the first conduction band and **b** hole-phonon scattering time in the valence band. Note the logarithmic scale for the false colour, showing the scattering time. The solid line depicts electron motion in the band along zigzag direction, the dotted line – along the armchair direction. Also, high symmetry points are marked. **c** Normalised harmonic yield as a function of the pump wavelength for the zigzag (top figure) and the armchair orientations (bottom figure), calculated by SBE with T2 = T(k) shown in panels a-b.

K (K') valley up to 1 ps and by a factor 2-3 across the band, remaining ≈3 fs near the Γ-point (see Supplementary Figure 6). The calculated ultrashort scattering time at low temperatures suggests that the processes of phonon emission in electron-lattice collisions play the dominant role in the dephasing time defining the HHG efficiency in solids. Also, the increased values of the scattering times at low temperatures suggest that the slope in the wavelength dependence of the harmonic yield should reduce at low temperatures, thus significantly increasing the harmonic yield. This, in turn, provides a possibility for a direct experimental determination of the contributions of the electron-electron and electron-phonon scatterings by conducting wavelength- and temperature-dependent HHG measurements for different laser intensities. It is worth noting that the dominant role of the electron-phonon scattering at relatively low densities of excited carriers was suggested previously in experiments on four-wave mixing in bulk semiconductors[36,37].

Simulations for a monolayer $MoS_2$ provide very similar results with slightly larger scattering time, that explains the difference in the slopes retrieved for $MoS_2$ and $WS_2$ monolayers. Finally, to show that electron-phonon scattering is the universal mechanism causing ultrafast dephasing in the electron motion outside the conduction band minima, we calculated the k-dependent total scattering rate in a bulk CdSe wurtzite crystal. Similar to the TMD monolayer, the scattering time drops abruptly from ≈100 fs at the Γ-point to a few femtoseconds when moving to larger lattice momenta (see Supplementary Fig.7).

To validate that electron-phonon scattering is responsible for the dephasing in HHG under the conditions of our experiments, the k-dependent scattering times calculated from DFT were incorporated into simulations of HHG in a $WS_2$ monolayer using the two-dimensional semiconductor Bloch equations (SBE) (see Methods and Supplementary Note 5 for details). Fig. 5c shows such calculated wavelength dependence of the harmonic yield for the laser pulse parameters and the spectral integration range used in the experiments. The retrieved slope of 7.3±0.2 (7.2±0.4) for the zigzag (armchair) orientation agrees well with the values retrieved from the experimental measurements. Also, the oscillating character of the HHG wavelength dependence, observed in the experiments, is well reproduced in the SBE simulations. Despite the strong dispersion of the electron-phonon scattering rate (see Fig.5 and Supplementary Note 6), as the electron motion covers large parts of the Brillouin zone, this leads to an efficient averaging of the dephasing time. Therefore, our experiments represent a new technique for all-optical measurements of an average electron-phonon scattering rate. The major contribution of our work is to finally clarify the physical origin of the ultrashort dephasing time, which until now has remained elusive and always been used simply as an empirical value. In this regard, we further stress that HHG has the peculiarity to probe an average electron-phonon scattering time over the entire Brillouin zone, in contrast to standard time-resolved (*e.g.*, pump-probe) measurements which only probe electron-phonon scattering in the proximity of the conduction band minimum (*e.g.*, the K valleys in TMDs).

## Discussion

In conclusion, we measured the wavelength dependence of the high-order harmonic yield in atomically thin (WS$_2$ and MoS$_2$) and bulk (CdSe) semiconductors. The measurements reveal a strong drop in the harmonic yield when increasing the excitation wavelength, significantly larger compared to the values retrieved for HHG in gases and with a material specific exponent in the range $\lambda^{-6}$ to $\lambda^{-9}$ that can only be reproduced by numerical simulations assuming an ultrashort dephasing time ≈4-5 fs. By analysing different possible scattering mechanisms that occur in solids, we suggest that electron-phonon scattering is the main source of decoherence in the electron-hole dynamics responsible for interband HHG. Using DFT calculations, we further show that this process is highly dispersive (k-dependent) and that in WS$_2$ it occurs on an ultrafast time scale down to 2 fs for electrons moving outside the K valleys. This ultrashort time scale is preserved also at low temperatures, suggesting that the phonon emission is the major scattering channel contributing to the decoherence time. From this we obtain excellent qualitative and quantitative agreement between the experimental results and simulations based on the ab-initio calculated electron-phonon scattering rate. Simulations for other materials, used in the experiments, including different crystal symmetry bulk CdSe, confirm that ultrafast electron-phonon scattering rate is a universal source of decoherence in solid HHG. Our work finally resolves the long lasting and open question about the role and physical origin of decoherence processes of carrier dynamics responsible for interband HHG. The strong k-dependence of phonon induced dephasing times resolves the contraction between ultra-high scattering rates present in HHG compared with comparably narrow line widths observed for low power excitations as e.g. excitons. Therefore, we propose wavelength-dependence measurements of the harmonic yield as an all-optical method to retrieve the average over the lattice momentum electron-phonon scattering rate. This result is thus paramount for the design of compact attosecond light sources based on HHG in solids and for applications of HHG in ultrafast time-resolved spectroscopy in condensed matter.

## Methods

### Experiments

A Ti:sapphire based regenerative amplifier (Coherent Astrella, 800 nm central wavelength, 35 fs pulse duration at 1 kHz repetition rate) was used to pump an optical parametric amplifier (Light Conversion TOPAS), followed by a noncolinear difference frequency generation (NDFG) module, offering ultrashort pulses with the duration from 60 to 100 fs in a spectral region from 3 to 12 μm. To ensure that the wavelength-dependent measurements were conducted under fixed peak intensity conditions, the pulses were characterised temporally, using the second harmonic frequency-resolved optical gating (SH-FROG) method, spatially, using a mid-IR CCD camera for measuring a focal intensity distribution. The dependence on the pump energy was measured for all wavelengths. The harmonics were measured in the 200 - 1100 nm wavelength range, limited by the spectrometer and the detector. All spectra were normalised on the camera and the grating sensitivity and recalculated in the frequency domain before yield integration. The resulting yield for each laser wavelength was interpolated as a function of the laser peak intensity. The final data were acquired from interpolation (see Supplementary Note 2). The experimental setup is shown in Supplementary figure S1. A laser beam was attenuated by a pair of wire-grid polarisers on a silicon substrate and focused by a 50 mm CaF2 lens at a normal incidence onto a TMD/CdSe sample deposited on a c-cut sapphire substrate. Pump polarisation was tuned by a broadband half-wave plate (B-Halle). Harmonic emission is collected by a $CaF_2$ f=25 mm lens in transmission geometry and measured by a Kymera-328i spectrometer equipped with a cooled UV-enhanced CCD camera. An on-site microscopy setup with an x50 objective lens (NUV Plan Apo 0.45 NA) was implemented for optical diagnostics and alignment of the samples. All experiments were performed at ambient conditions.

## Numerical simulations

The geometry and the electronic structure of single-layer $WS_2$ was calculated using density functional theory as implemented in Quantum Espresso[38]. Phonon properties were obtained within density functional perturbation theory using the same software. For all calculations, we used the SG15 optimised norm-conserving pseudo-potentials[39,40] from http://www.quantum-simulation.org/potentials/sg15_oncv/index.htm. A kinetic energy cutoff for the wavefunction of 90 Ry was applied and the PBE functional was used in all cases[41]. To ensure a sufficient decoupling of adjacent layers, we added about 19.4 Å of vacuum in the direction perpendicular to the layer. For the electron and phonon calculations we used Monkhorst-Pack meshes with 18x18x1 and 9x9x1 points, respectively. The electronic band-structure and the phonon dispersions are shown in Supplementary Figure 5. The electron-phonon induced finite lifetime of the electronic states was obtained using the electron-phonon Wannier functions (EPW) approach[42,43]. It is calculated from the imaginary part of the electron-phonon self-energy for the respective Bloch state as in (Ref. 44). For the integration over the Brillouin zone, we used 27x27x1 *k*-points and a very fine mesh of 500x500x1 *q*-points to assure convergence.

We carried out the rt-TDDFT calculations using the Octopus code[45] with the "hgh_lda" pseudopotentials, the semicore set of Hartwigsen-Goedecker-Hutter in the local density approximation[46]. The simulations converged for a Brillouin zone sampling of 40×40 and a real-space mesh with 0.3 Bohr spacing.

For the SBE simulations, we followed the approach outlined in (Ref. 2). To calculate the energy bands and transition dipole elements as well as Berry connections, we used a three-band tight-binding model[47] including one valence and two conduction bands and spin-orbit coupling. The details of SBE model are elaborated in the Supplementary Note 4.

## Acknowledgments

This work was funded by the German Research Foundation DFG (CRC 1375 NOA), project number 398816777 (subproject C4, A1), the Federal Ministry of Education and Research (BMBF) under the "Make our Planet Great Again – German Research Initiative" (grant 57427209) implemented by the German Academic Exchange Service (DAAD)

## Authors contribution

V.Ko. build the experimental setup, carried out experimental measurements and data processing, T. L. conducted rt-TDDFT and SBE numerical simulations, V.Kr. prepared and characterised the monolayers, A.C. conducted DFT calculations for band structure and electron-phonon scattering rates, M. Z. contributed by the laser system, results discussion, and interpretation, C.S. supervised the research work, M.W. synthesised the CdSe film, U.P. and S.G. supervised the numerical simulations, contributed to results analysis and interpretation, G.S. supervised monolayer manufacturing, contributed to results analysis and interpretation, D.K. conceived the idea, supervised the experimental measurements, contributed to data processing and interpretation. V.Ko. and D.K. prepared the manuscript with contribution from all co-authors.

† These authors contributed equally

**References**

1. Vampa, G., Brabec, T. Merge of high harmonic generation from gases and solids and its implications for attosecond science. *Journ. Phys. B* **50**, 083001 (2017).
2. Yue, L., Gaarde, M. B., Introduction to theory of high-harmonic generation in solids: tutorial. *Journ. Phys. B* **39**, 535-555 (2022).
3. Gouliemakis, E., Brabec, T. High-harmonic generation in condensed matter. *Nature Phot.* **16**, 411–421 (2022).
4. Luu, T.T., Garg, M., Kruchinin, S. Yu., Moulet, A., Hassan, M. Th., Goulielmakis, E. Extreme ultraviolet high-harmonic spectroscopy of solids. *Nature* **521**, 498-502 (2015).
5. Vampa, G., Hammond, T. J., Thiré, N., Schmidt, B. E., Légaré, F., McDonald, C. R., Brabec, T., Klug, D. D., Corkum, P. B. All-optical reconstruction of crystal band structure. *Phys. Rev. Lett.* **115**, 193603 (2015).
6. Lanin, A. A., Stepanov, E. A., Fedotov, A. B., Zheltikov, A. M. Mapping the electron band structure by intraband high-harmonic generation in solids. *Optica* **4**, 516-519 (2017).
7. Yoshikawa, N., Tamaya, T., Tanaka, K. High-harmonic generation in graphene enhanced by elliptically polarized light excitation. *Science* **356**, 736–738 (2017).
8. Kobayashi, Y., Heide, C., Kelardeh, H. K., Johnson, A., Liu, F., Heinz, T. F., Reis, D. A., Ghimire, S. Polarisation flipping of even-order harmonics in monolayer transition-metal dichalcogenides. *Ultrafast Sci.* **2021**, 1–9 (2021).
9. Liu, H., Li, Y., You, Y. S., Ghimire, S., Heinz, T. F., Reis, D. A. High harmonic generation from an atomically thin semiconductor. *Nature. Phys.* **13**, 262–265 (2017).
10. Silva, R. E., Jiménez-Galán, A., Amorim, B., Smirnova, O., Ivanov, M. Topological strong-field physics on sub-laser-cycle timescale. *Nature. Phot.* **13**, 849–854 (2019).
11. Chacón, A., Kim, D., Zhu, W., Kelly, S. P., Dauphin, A., Pisanty, E., Maxwell, A. S., Picón, A., Ciappina, M. F., Kim, D. E., Ticknor, C., Saxena, A., Lewenstein, M. Circular dichroism in higher-order harmonic generation: heralding topological phases and transitions in Chern insulators. *Phys. Rev. B* **102**, 134115 (2020).

12. Bai, Y., Fei, F., Wang, S., Li, N., Li, X., Song, F., Li, R., Xu, Z., Liu, P. High harmonic generation from topological surface states. *Nature Phys*. **17**, 311–315 (2020).
13. Baykusheva, D., Chacón, A., Kim, D., Kim, D. E., Reis, D. A., Ghimire, S. Strong-field physics in three-dimensional topological insulators. *Phys. Rev. A* **103**, 023101 (2021).
14. Schmid, C. P., Weigl, L., Grössing, P., Junk, V., Gorini, C., Schlauderer, S., Ito, S., Meierhofer, M., Hofmann, N., Afanasiev, D., Crewse, J., Kokh, K. A., Tereshchenko, O. E., Güdde, J., Evers, F., Wilhelm, J., Richter, K., Höfer, U., Huber, R. Tunable non-integer high-harmonic generation in a topological insulator. *Nature* **593**, 385–390 (2021).
15. Golde, D., Meier, T., Koch, S. W. High harmonics generated in semiconductor nanostructures by the coupled dynamics of optical inter- and intraband excitations. *Phys. Rev. B* **77**, 075330 (2008).
16. Vampa, G., McDonald, C. R., Orlando, G., Klug, D. D., Corkum, P. B., Brabec, T. Theoretical Analysis of High-Harmonic Generation in Solids. *Phys. Rev. Lett*. **113**, 073901 (2014).
17. Herrmann, P., Klimmer, S., Lettau, T., Monfared, M., Staude, I., Paradisanos, I., Peschel, U. Soavi, G. Nonlinear All-Optical Coherent Generation and Read-Out of Valleys in Atomically Thin Semiconductors. *Small* **19**, 2301126 (2023).
18. Floss, I., Lemell, C., Wachter, G., Smejkal, V., Sato, S. A., Tong, X.-M., Yabana, K., Burgdörfer, J. *Ab initio* multiscale simulation of high-order harmonic generation in solids. *Phys. Rev. A* **97**, 011401(R) (2018).
19. Liu, X., Li, L., Zhu, X., Huang, T., Zhang, X., Wang, D., Lan, P., Lu, P. Wavelength dependence of high-order harmonic yields in solids. *Phys. Rev. A* **98**, 063419 (2018).
20. Wang, Z., Park, H., Lai, Y. H., Xu, J., Blaga, C. I., Yang, F., Agostini, P., DiMauro, L. F. The roles of photo-carrier doping and driving wavelength in high harmonic generation from a semiconductor. *Nature Comm*. Doi:10.1038/s41467-017-01899-1 (2017).
21. Brown, G. G., Jiménez-Galán, A., Silva, R. E. F., Ivanov, M. A real-space perspective on dephasing in solid state high harmonic generation. https://arxiv.org/abs/2210.16889
22. Brown, G. G., Jiménez-Galán, A., Silva, R. E. F., Ivanov, M. Ultrafast dephasing in solid state high harmonic generation: microscopic origin revealed by real-space dynamics. https://arxiv.org/abs/2310.17005
23. Ninomiya, S., Adachi, S., Optical properties of cubic and hexagonal CdSe. *Journ. Appl. Phys*. **78**, 4681-4689 (1995).

24. Schiessl, K., Ishikawa, K. L., Persson, E., Burgdörfer, J. Quantum Path Interference in the Wavelength Dependence of High-Harmonic Generation. *Phys. Rev. Lett.* **99**, 253903 (2007).
25. Dogadov, O., Trovatello, C., Yao, B., Soavi, G., Cerullo, G. Parametric nonlinear optics with layered materials and related heterostructures. *Laser Phot. Rev.* **16**, 210072 (2022).
26. Barati, F., Grossnickle, M., Su, S., Lake, R. K., Aji, V., Gabor, N. M. Hot carrier-enhanced interlayer electron–hole pair multiplication in 2D semiconductor heterostructure photocells. *Nature Nanotech.* **12**, 1134-1139 (2017).
27. Kim, Ji-H., Bergren, M. R., Park, J. C., Adhikari, S., Lorke, M., Frauenheim, T., Choe, D.-H., Kim, B., Choi, H., Gregorkiewicz, T., Lee, Y. H. Carrier multiplication in van der Waals layered transition metal dichalcogenides. *Nature Commun.* https://doi.org/10.1038/s41467-019-13325-9 (2019).
28. Zheng, W., Bonn, M., Wang, H. I. Photoconductivity multiplication in semiconducting few-layer $MoTe_2$. *Nano Lett.* **20**, 5807-5813 (2020).
29. He, G., Nathawat, J., Kwan, C.-P., Ramamoorthy, H., Somphonsane, R., Zhao, M., Ghosh, K., Singisetti, U., Perea-López, N., Zhou, C., Elías, A. L., Terrones, M., Gong, Y., Zhang, X., Vajtai, R., Ajayan, P. M., Ferry, D. K., Bird, J. P. Negative differential conductance & hot-carrier avalanching in monolayer $WS_2$ FETs. *Nature Scientific Rep.* DOI:10.1038/s41598-017-11647-6 (2017).
30. D. K. Ferry. Electron transport in some transition metal dichalcogenides: $MoS_2$ and $WS_2$. *Semicond. Sci. Technol.* **32**, 085003 (2017).
31. Keldysh, L. V. Ionization in the field of a strong electromagnetic wave. *Sov. Phys. JETP* **20**, 1307-1314 (1965).
32. Laser induced damedge in optical materials. Edited by D. Ristau, CRC Press, ISBN 13: 978-1-138-19956-9, 2016.
33. Heide, C., Kobayashi, Y., Johnson, A. C., Liu, F., Heinz, T. F., Reis, D. A., Ghimire, S. Probing electron-hole coherence in strongly driven 2D materials using high-harmonic generation. *Optica* **9**, 512-516 (2022).
34. Jin, Z., Li, X., Mullen, J. T., Kim, K. W. Intrinsic transport properties of electrons and holes in monolayer transition-metal dichalcogenides. *Phys. Rev. B* **90**, 045422 (2014).
35. Sohier, T., Campi, D., Marzari, N., Gibertini, M. Mobility of two-dimensional materials from first principles in an accurate and automated framework. *Phys. Rev. Materials* **2**, 114010 (2018).
36. Arlt, S., Siegner, U., Kunde, J., Morier-Genoud, F., Keller, U. Ultrafast dephasing of continuum transitions in bulk semiconductors. *Phys. Rev. B* **59**, 14860 (1999).
37. Becker, P. C., Fragnito, H. L., Brito Cruz, C. H., Fork, R. L., Cunningham, J. E., Henry, J. E., Shank, C. U. Femtosecond photon echoes from band-to-band transitions in GaAs. *Phys. Rev. Lett.* **61**, 1647-1649 (1988).

38. Giannozzi, P., Baseggio, O., Bonfà, P., Brunato, D., Car, R., Carnimeo, I., Cavazzoni, C., de Gironcoli, S., Delugas, P., Ruffino, F. F., Ferretti, A., Marzari, N., Timrov, I., Urru, A., Baroni, S. QUANTUM ESPRESSO toward the exascale. *Jorn. Chem. Phys.* **152**, 154105 (2020); http://www.quantum-espresso.org
39. Hamann, D. R. Optimized norm-conserving Vanderbilt pseudopotentials. *Phys. Rev. B* **88**, 085117 (2013).
40. Schlipf, M., Gygi, F. Optimization algorithm for the generation of ONCV pseudopotentials. *Comp. Phys. Commun.* **196**, 36-44 (2015).
41. Perdew, J. P., Burke, K., Ernzerhof, M. Generalized Gradient Approximation Made Simple. *Phys. Rev. Lett.* **77**, 3865-3868 (1996).
42. Giustino, F., Cohen, M. L., Louie, S. G. Electron-phonon interaction using Wannier functions. *Phys. Rev. B* **76**, 165108 (2007).
43. Poncé, S., Margine, E. R., Verdi, C., Giustino, F. EPW: Electron–phonon coupling, transport and superconducting properties using maximally localized Wannier functions. *Comput. Phys. Commun.* **209**, 116-133 (2016).
44. Mahatha, S. K., Ngankeu, A. S., Frank Hinsche, N., Mertig, I., Guilloy, K., Matzen, P. L., Bianchi, M., Sanders, C. E., Miwa, J. A., Bana, H., Travaglia, E., Lacovig, P., Bignardi, L., Lizzit, D., Larciprete, R., Baraldi, A., Lizzit, S., Hofmann, P. Electron–phonon coupling in single-layer $MoS_2$. *Surf. Sci.* **681**, 64-69 (2019).
45. Tancogne-Dejean, N., Oliveira, M. J. T., Andrade, X., Appel, H., Borca, C. H., Le Breton, G., Buchholz, F., Castro, A., Corni, S., Correa, A. A., De Giovannini, U., Delgado, A., Eich, F. G., Flick, J., Gil, G., Gomez, A., Helbig, N., Hübener, H., Jestädt, R., Jornet-Somoza, J., Larsen, A. H., Lebedeva, I. V., Lüders, M., Marques, M. A., L., Ohlmann, S. T., Pipolo, S., Rampp, M., Rozzi, C. A., Strubbe, D. A., Sato, S. A., Schäfer, C., Theophilou, I., Welden, A., Rubio, A. Octopus, a computational framework for exploring light-driven phenomena and quantum dynamics in extended and finite systems. *Journ. Chem. Phys.* **152** 124119 (2020).
46. Hartwigsen, C., Goedecker, S., Hutter, J. Relativistic separable dual-space Gaussian pseudopotentials from H to Rn. *Phys. Rev. B* **58**, 3641-3662 (1998).
47. Liu, G.-B., Shan, W.-Y., Yao, Y., Yao, W., Xiao, D. Three-band tight-binding model for monolayers of group-VIB transition metal dichalcogenides. *Phys. Rev. B* **88**, 085433 (2013).

# Supplementary Information

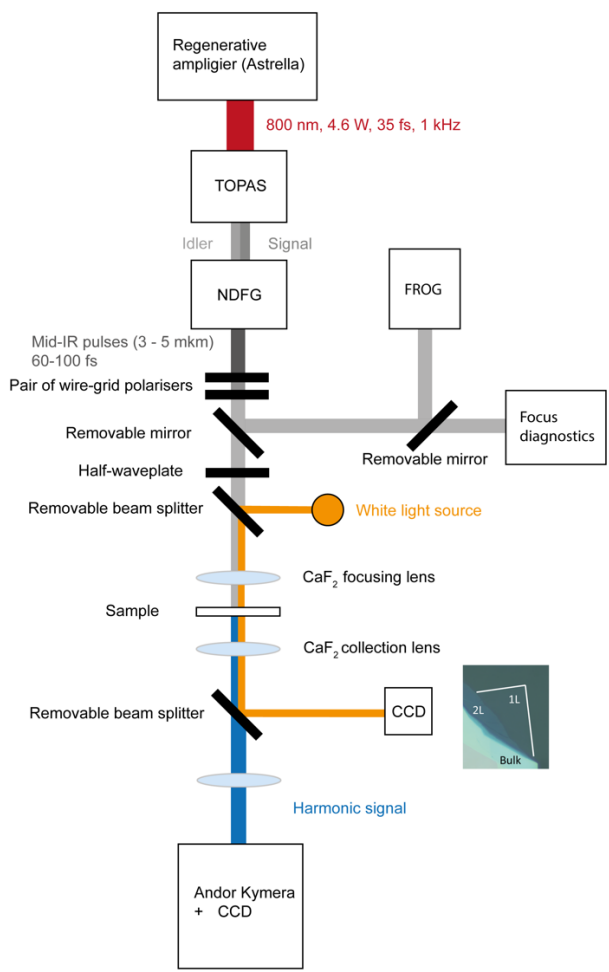

Supplementary Figure 1. Schematic presentation of the experimental setup.

**Supplementary Note 1: Sample Fabrication and Characterisation**

$MoS_2$ and $WS_2$ monolayer samples were prepared by mechanical exfoliation from a commercial bulk synthetic crystal (HQGraphene) using an adhesive tape (Bluetape Minitron) and subsequently transferred to sapphire substrates by the PDMS (Gel-Pak) dry transfer method [1]. The samples were identified by optical contrast and their monolayer nature was confirmed by micro-photoluminescence spectroscopy in reflection geometry from a $SiO_2$/Si substrate. PL measurements were performed with CW excitation (Cobolt 08-DPL 532nm) using a 50x objective with a numerical aperture of 0.42 (Mitutoyo Plan Apo) resulting in a focal spot size of ~1µm. We observed PL maxima at ~ 616nm (~2.01 eV) / 629nm (~1.97 eV) corresponding to the A-exciton/trion of $WS_2$ (Supplementary Figure 2a), and at ~ 652nm (~1.9 eV) / 602nm (~2.05 eV) / 665nm (~1.86 eV) corresponding respectively to the A-exciton, B- exciton and trion of $MoS_2$ (Supplementary Figure 2b), in agreement with literature [2,3].

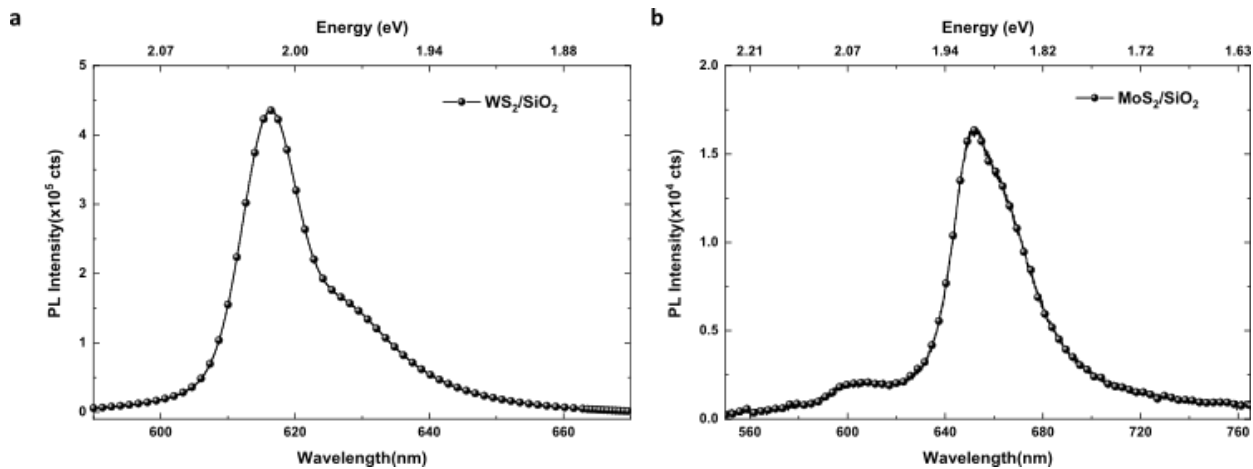

Supplementary Figure 2. Photoluminescence characterisation of monolayer (a) $WS_2$ and (b) $MoS_2$ on $SiO_2$/Si substrates.

## Supplementary Note 2: Fixing the Laser Intensity for all Wavelengths

To analyse the wavelength dependence, it is essential to keep the peak intensity of the laser pulse fixed. To achieve this, we first performed harmonic-yield measurements as a function of an average power together with FROG traces and focal intensity distributions (Supplementary Figure 3a and c) for every wavelength in the experiment. To precisely characterise the focal intensity, aside from the main setup, we installed a long-focusing CaF2 lens (200 mm) on a translation stage and measured with a MID-IR camera the spot size as a function of the lens position. We attributed the minimal spot size to the maximum harmonic output and, assuming negligible chromatic aberrations, recalculated the focal intensity to the 50 mm $CaF_2$ lens used in the experiment. When the pulse is characterised in space and time domain, we can recalculate the harmonic yield as a function of peak intensity and, presuming a smooth dependence, fit the results with a polynomial (Supplementary Figure 3c and d). Fixing the intensity for one of the laser wavelengths, we interpolated the harmonic yield for all other wavelengths, using the retrieved polynomial fits, as it is shown in Supplementary Figure 3.

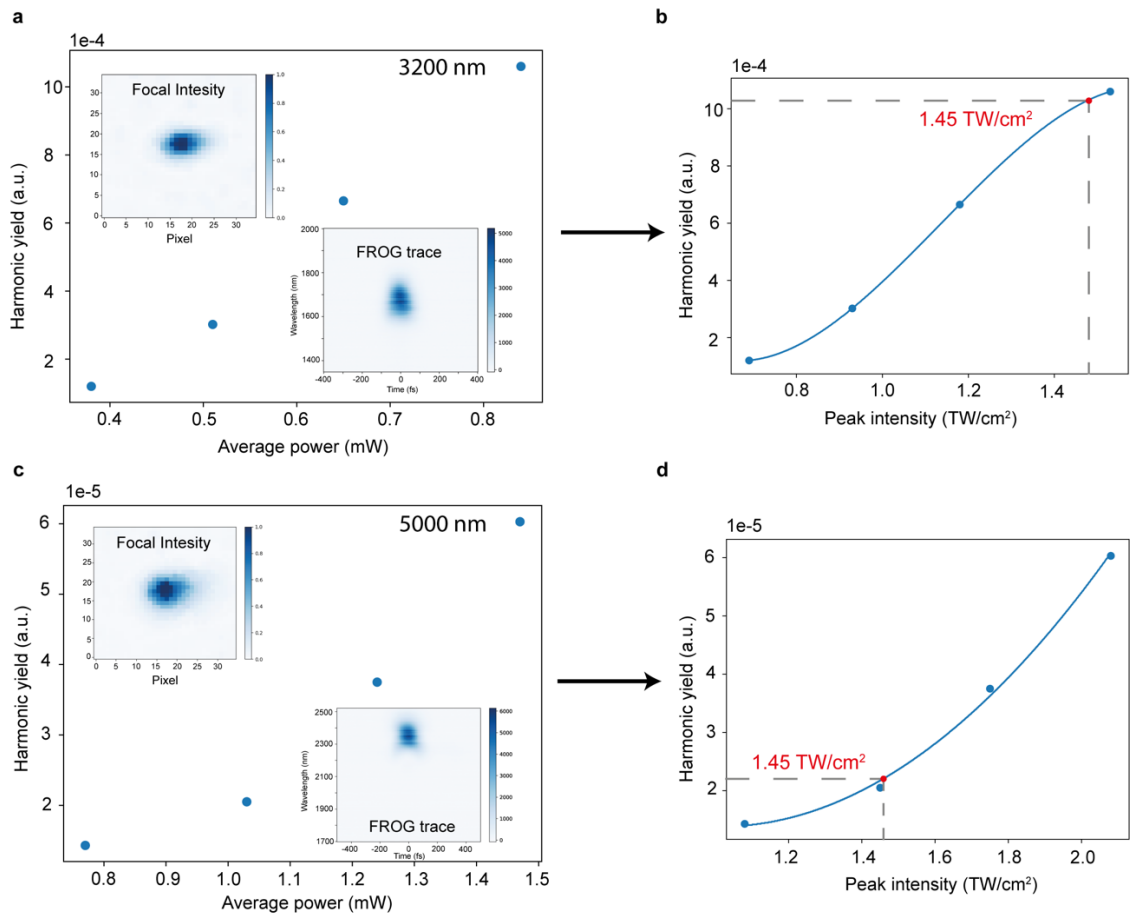

Supplementary Figure 3. Calibration of laser intensity. **a** Harmonic yield as a function of a laser average power, measured in the experiment, inset figures represent a 2D focal intensity distribution and FROG trace at 3200 nm central wavelength. **b** Harmonic yield as a function of peak laser intensity, recalculated according to FROG traces and focal distribution. The solid line shows a polynomial fit, and the dashed grey line represents the choice of desired intensity analysed in the manuscript. **c** and **d** Same but for 5000 nm central wavelength.

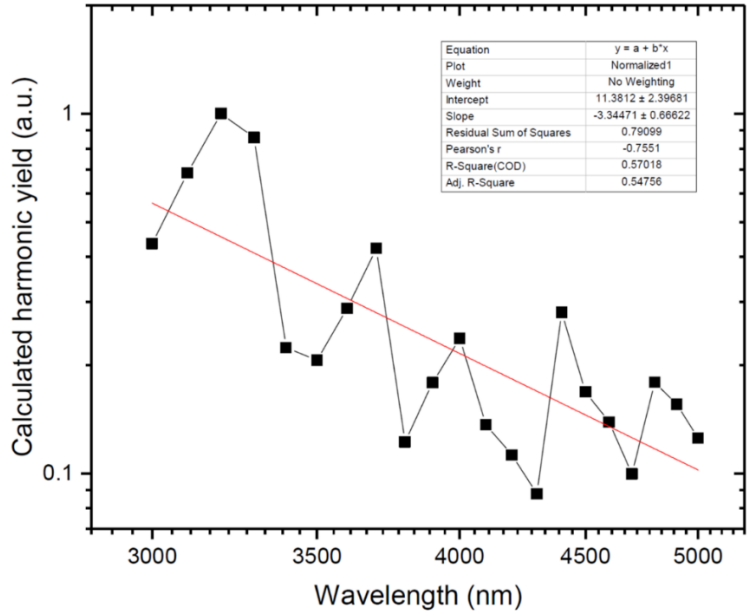

Supplementary Figure 4. rt-TDDTF simulations. The wavelength dependence in the harmonic yield within the spectral range measured in the experiments.

## Supplementary Note 3: Electron-Phonon Scattering Rates

The optimised lattice constant was found to be $a = 3.188\ Å$ in agreement with previously reported results [4, 5]. The calculated bandgap is $E_g = 1.86\ eV$ and therefore smaller than the experimentally observed value ($\approx 2\ eV$). Supplementary Figure 5a shows the resulting electronic band-structure $\varepsilon_n(\vec{k})$ for a selection of bands around the Fermi energy. The phonon dispersion $\omega_\lambda(\vec{q})$ in Supplementary Figure 5b displays the quadratic dispersion of the ZA mode which is characteristic for 2D materials.

To verify the quality of the Wannierisation procedure which is needed for the computation of the electron-phonon self-energy, we compare the band-structure and the phonon dispersions obtained with electron-phonon Wannier functions (EPW) to the respective results from Quantum Espresso (QE) [Ref.38] simulations. As follows from Supplementary Figure 5, there is a very good agreement between the original (QE) and the interpolated data (EPW).

The electron-phonon scattering rate $1/\tau_n(\vec{k})$ is obtained from the first order electron-phonon self-energy as follows [43]:

$$\frac{1}{\tau_n(\vec{k})} = \frac{2\pi}{\hbar} \sum_{n,\vec{q},\lambda} |g_{nm}^\lambda(\vec{k},\vec{q})|^2 \{[n_\lambda^0(\vec{q}) + f_n^0(\vec{k}+\vec{q})]\delta(\varepsilon_n(\vec{k}+\vec{q}) - \hbar\omega_\lambda(\vec{q}) - \varepsilon_m(\vec{k}))$$
$$+ [n_\lambda^0(\vec{q}) + 1 - f_n^0(\vec{k}+\vec{q})]\delta(\varepsilon_n(\vec{k}+\vec{q}) + \hbar\omega_\lambda(\vec{q}) - \varepsilon_m(\vec{k}))\}.$$

Here, $g_{nm}^\lambda(\vec{k},\vec{q})$ is the electron-phonon scattering matrix element, $f_n^0(\vec{k})$ and $n_\lambda^0(\vec{q})$ denote the Fermi and Bose distributions for electrons and phonons, respectively. The electron-phonon scattering time $\tau_n(\vec{k})$ for the first conduction and the highest valence bands is shown in Supplementary Figure 6 for temperature 20 K. To demonstrate the ultrafast nature of the electron-phonon scattering in solids, Supplementary Figure 7 shows the scattering rate along main symmetry directions in a wurtzite CdSe crystal.

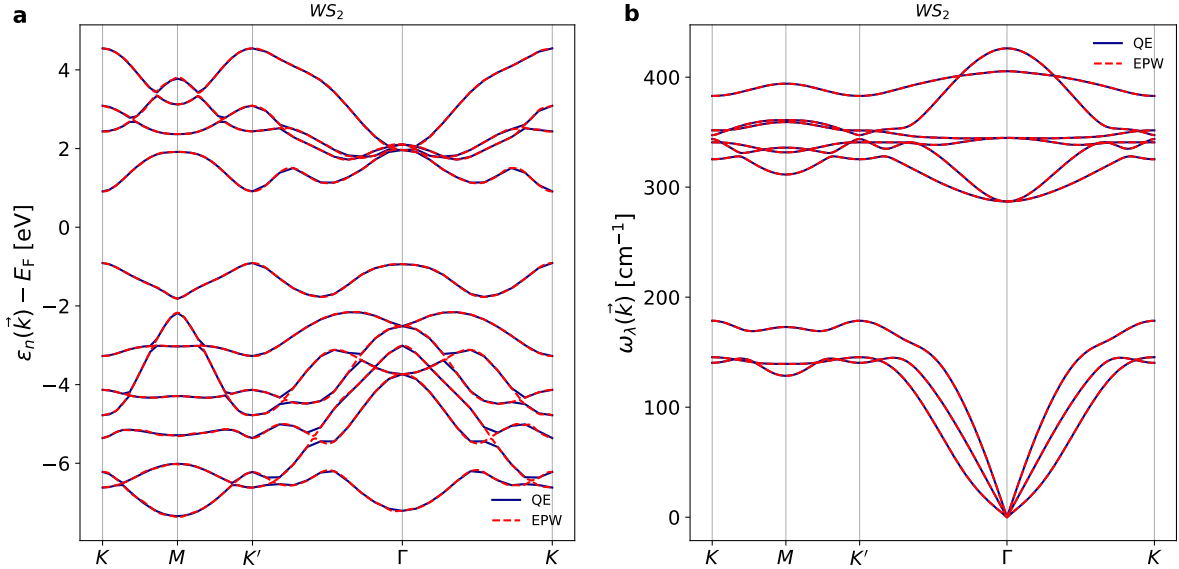

Supplementary Figure 5. Comparison of **a**) the electronic band-structure and **b**) the phonon dispersions calculated with Quantum espresso (QE, blue solid lines) and EPW (red dashed lines).

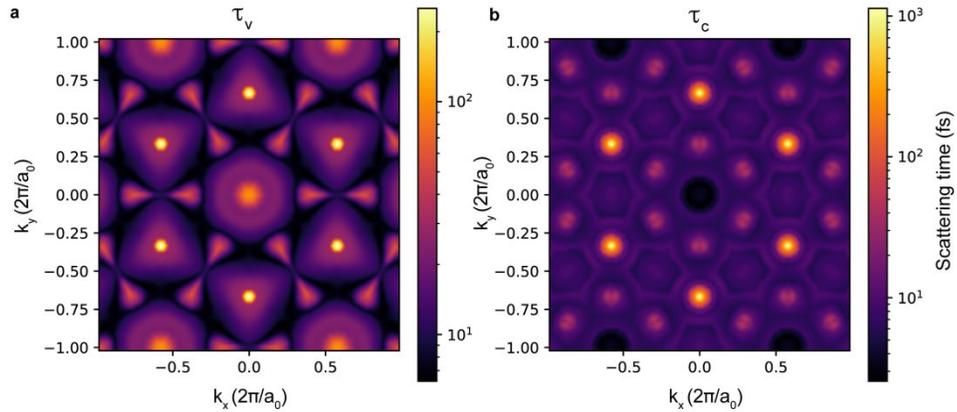

Supplementary Figure 6. Temperature dependence of the electron-phonon scattering. k-dependent electron-phonon scattering time for the upper valence band **a**) at the temperature 20 K and for the first conduction band **b**) at the temperature 20 K. Note the logarithmic scale for the plots.

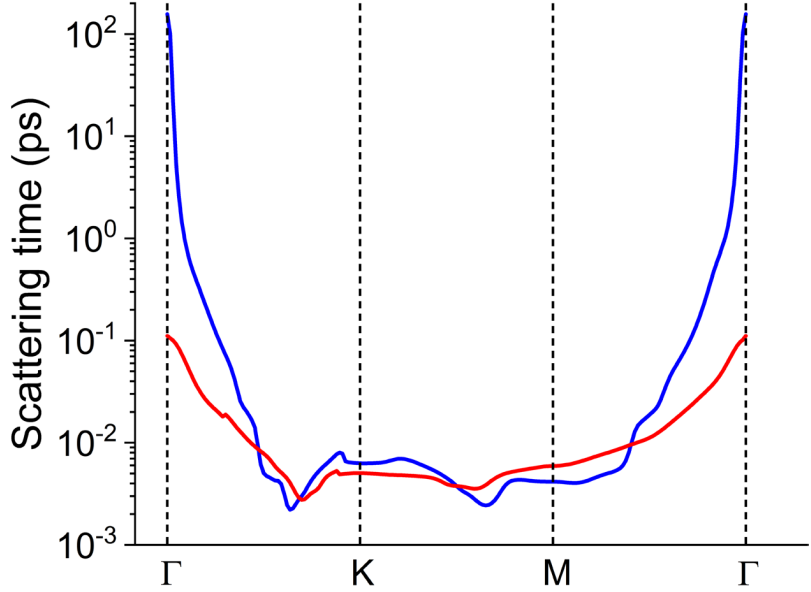

Supplementary Figure 7. k-dependent electron-phonon scattering time for the upper valence band (red line) and the lowest conduction band (blue line) in wurtzite CdSe crystal at 293 K temperature.

## Supplementary Note 4: HHG Simulations with Semiconductor Bloch Equations

In our simulations of HHG with the SBE approach we consider the upper valence and the first conduction bands. To include electron-phonon scattering processes in the SBE equations, we assume that the scattering is primarily affecting the coherency between carriers. Therefore, we assume that every scattering process is influencing the dephasing of the polarisation and do not consider the change in the population at a specific lattice momentum caused by carrier loss and influx due to the scattering. Thus, the dephasing is included in the SBE equation for the polarisation only, where each carrier involved is adding one half of its scattering rate (see e.g. [6]). The inverse of the dephasing time for the polarisation $T_{m,n}(k)$ between two bands $m$ and $n$ is calculated as the average of the inverse scattering times, i.e. scattering rates, retrieved from the electron-phonon scattering calculations:

$$\frac{1}{T_{m,n}(k)} = \frac{1}{2}\left(\frac{1}{\tau_m(k)} + \frac{1}{\tau_n(k)}\right).$$

Thus, the dephasing time is dominated by the shortest involved scattering time. Supplementary Figure 8 shows the corresponding dephasing times affecting the polarisation between different bands.

The system of the SBE equations is obtained by plugging the Bloch ansatz $\sum_{m,k\in BZ} a_m^k(t)\, \varphi_m^k(r)$ into the minimal coupling Hamiltonian $\mathcal{H}$ in length gauge and solving for the time dependent coefficients $a_m^k(t)$. Here, $\varphi_m^k$ is the Bloch function for band index $m$, with the associated energy band $\mathcal{H}\varphi_m^k = \varepsilon_m^k \varphi_m^k$. If we denote the products of two

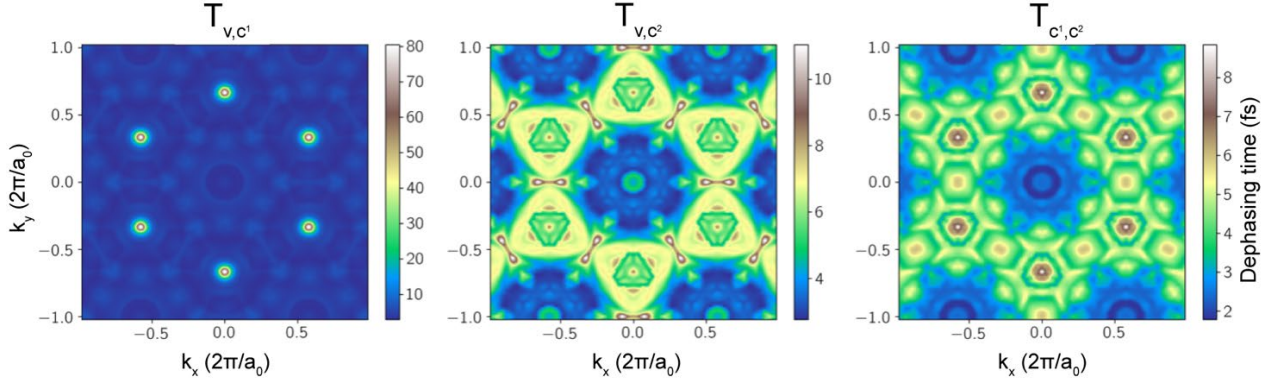

Supplementary Figure 8. k-dependent dephasing times used in the SBE simulations for the dephasing between a) the valence and the first conduction band b) the valence ansd the second conduction band c) the two conduction bands.

coefficients as $\rho_{mn}^k = a_m^k(t)a_n^{k*}(t)$, the equations read

$$i\dot{\rho}_{mn}^k(t) = \left(\varepsilon_m^k - \varepsilon_m^k - \frac{1-\delta_{mn}}{T_m^k}\right)\rho_{mn}^k(t)$$
$$+ E(t) \cdot \sum_l [D_{ml}^k \rho_{ln}^k(t) - D_{ln}^k \rho_{ml}^k(t)]$$
$$+ iE(t) \cdot \nabla_k \rho_{mn}^k(t)$$

where $E(t)$ is the electrical field and $D_{mn}^k$ are the transition dipole elements for $m \neq n$ and the Berry connections for $m = n$. They are calculated by $D_{mn}^k = \langle \varphi_m^k | \nabla_k | \varphi_n^k \rangle$ and the scalar product is taken with respect to the unit cell. The last term in the equations, which couples the different k-points, can be omitted by transforming into a comoving frame and introducing a time dependent $k = K(t) + A(t)$. In that case, the quantity $D_{mn}^k \to D_{mn}^{K+A(t)}$ becomes time dependent and is shifted, where the magnitude of the shift is given by the vector potential $A(t)$. The first line contains the dephasing time $T_m^k$ which in our case depends on the band index $m$ and $k$. A concise overview of different approaches for the SBE can be found in [Ref.2].

The band structures and dipole elements were calculated using a tight binding model developed in [Ref.47]. In these calculations we included spin-orbit coupling resulting in fully non-degenerate bands. The simulations were converged with a maximal time step of $\Delta t = 0.2$ a.u. and a Brillouin zone sampling of $300 \times 300$ points. The maximal field amplitude was set to $1.67 \cdot 10^{-3}$ a.u.

## Supplementary Note 5: Comparison Between Calculations with k-dependent and k-independent Dephasing Time

Despite the strong k-dependence of the electron-phonon scattering rate (see Supplementary Figure 6), electron/hole motion in the bands covering a large range of the lattice momentum results in an effective averaging of the dephasing over the Brillouin zone. Therefore, we compare our numerical results based on the ab-initio k-dependent electron-phonon scattering time to the most commonly used approach for HHG in solids which is considering a constant, k-independent ultrashort dephasing time. The summary of such simulations is shown in Supplementary Figure 9 together with the slope values retrieved from the experimental data. The results of calculations with k-independent dephasing time suggest, that the dependence of the slope upon the value of the dephasing time can be very well approximated (R-square is 0.9987) by the exponential function $e^{-\tau/\Delta\tau}$ with the decrement $\Delta\tau \approx 2.76 \pm 0.29$ (the dashed line in Supplementary Figure 9). These results are in a good agreement with the results previously calculated for bulk ZnO crystal and reported in [Ref.19]. A comparison between the simulations with k-dependent dephasing time and k-independent constant dephasing time suggests, that for calculations of the integral harmonic yield, the former can be approximated by an average value that can be interpreted as an effective, material specific electron-phonon scattering rate.

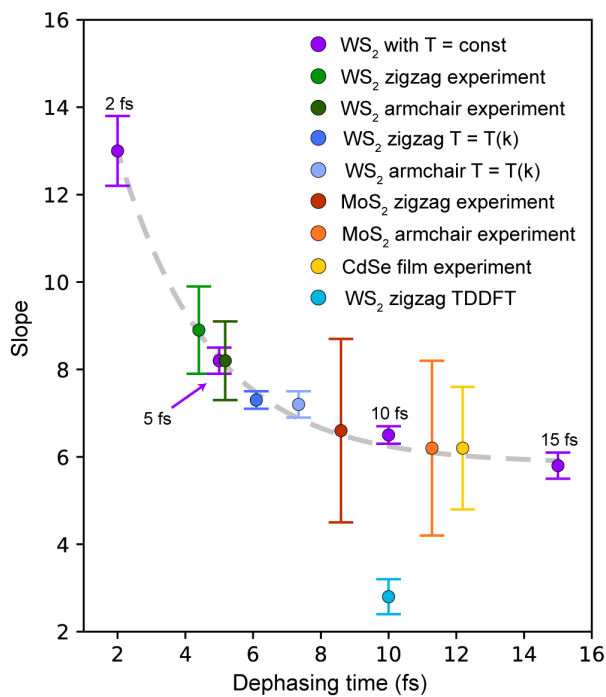

Supplementary Figure 9. Summary of the numerical simulations. The dashed line shows the slope value dependence on the dephasing time in calculations when the latter is a k-independent constant. The slope values retrieved from the experimental measurements and k-dependent simulations are superimposed with constant dephasing time calculations for comparison.

References

[1] Frisenda, R. *et al.* Recent progress in the assembly of nanodevices and van der Waals heterostructures by deterministic placement of 2D materials. *Chem Soc Rev* **47**, 53–68 (2018).

[2] Mouri, S., Miyauchi, Y. & Matsuda, K. Tunable Photoluminescence of Monolayer MoS$_2$ via Chemical Doping. *Nano Lett* **13**, 5944–5948 (2013).

[3] Plechinger, G. *et al.* Identification of excitons, trions and biexcitons in single-layer WS$_2$. *physica status solidi (RRL) - Rapid Research Letters* **9**, 457–461 (2015).

[4] S. Haastrup, M. Strange, M. Pandey, T. Deilmann, P. S. Schmidt, N. F Hinsche, M. N Gjerding, D. Torelli, P. M Larsen, A. C Riis-Jensen. The Computational 2D Materials Database: high-throughput modeling and discovery of atomically thin crystals. 2D Mater. 5, 042002 (2018).

[5] M. N. Gjerding, A. Taghizadeh, A. Rasmussen, S. Ali, F. Bertoldo, T. Deilmann, N. R. Knøsgaard, M. Kruse, A. H. Larsen, S. Manti. Recent progress of the Computational 2D Materials Database (C2DB). 2D Mater. 8, 044002 (2021).

[6] T. Lettau, H.A.M Leymann & J. Wiersig. Pitfalls in the theory of carrier dynamics in semiconductor quantum dots: Single-particle basis versus the many-particle configuration basis. Physical Review B, 95(8) (2017).